\documentstyle[preprint,aps,pre,eqsecnum]{revtex}
\draft
\begin{document}

\title{Classical Infinite-Range-Interaction Heisenberg Ferromagnetic
\\
Model: Metastability and Sensitivity to Initial Conditions}

\vskip \baselineskip

\author{Fernando D. Nobre$^{1,2}$\thanks{Corresponding author:
E-mail address: nobre@dfte.ufrn.br} and 
Constantino Tsallis$^{1}$\thanks{E-mail address: tsallis@cbpf.br}}


\address{
$^{1}$Centro Brasileiro de Pesquisas F\'{\i}sicas \\
Rua Xavier Sigaud 150 \\
22290-180 \hspace{5mm} Rio de Janeiro - RJ \hspace{5mm} Brazil}

\vspace{1cm}

\address{
$^{2}$Departamento de F\'{\i}sica Te\'orica e Experimental \\
Universidade Federal do Rio Grande do Norte \\
Campus Universit\'{a}rio -- Caixa Postal 1641 \\
59072-970 \hspace{5mm} Natal - RN \hspace{5mm} Brazil}

\date{\today}
\maketitle

\vskip \baselineskip

\begin{abstract}
A N-sized inertial classical Heisenberg ferromagnet, which consists in a 
modification of the well-known standard model, where the spins 
are replaced by classical 
rotators, is studied in the limit of infinite-range interactions. The 
usual canonical-ensemble mean-field solution of the inertial classical 
$n$-vector ferromagnet (for which $n=3$ recovers the particular Heisenberg 
model considered herein) is briefly reviewed, showing  
the well-known second-order phase transition. 
This Heisenberg model is studied numerically 
within the microcanonical ensemble, through molecular dynamics. 
In what concerns the caloric curve, 
it is shown that, far from criticality, the kinetic temperature 
obtained at the long-time-limit microcanonical-ensemble simulation recovers
well the equilibrium canonical-ensemble estimate, whereas, close to 
criticality, a discrepancy (presumably due to finite-size effects) is found. 
The time evolution of the kinetic
temperature indicates that a basin of attraction exists for the
initial conditions for which
the system evolves into a metastable state, whose duration
diverges as $N \rightarrow \infty$, before attaining the 
terminal thermal equilibrium.
Such a metastable state is observed for a whole range of energies, which
starts right below criticality and extends up to very high energies (in fact,
the gap between the kinetic temperatures associated with the metastable
and the terminal equilibrium states is expected to disappear only as
one approaches infinite energy). To our knowledge, this has never
before been observed on similar Hamiltonian models, 
in a noticeable way, for such a large
range of energies. For example, for the XY ($n=2$) version of the present model,
such behavior was observed only near criticality.
It is shown also that the (metastable state) 
maximum Lyapunov exponent decreases with $N$ like 
$\lambda_{\rm max} \sim N^{-\kappa}$, where  
for the initial conditions
employed herein (maximal magnetization), $\kappa=0.225 \pm 0.030$,
both above and below the critical point. 

\vskip \baselineskip


\end{abstract}

\newpage

\noindent
{\large\bf 1. \quad Introduction}

\vskip \baselineskip

The Boltzmann-Gibbs (BG) statistical-mechanics represents one of
the most successful physical theories, providing a good description of 
many experimental systems at equilibrium
\cite{pathria,stanley,landsberg}. The applicability of such a formalism is 
justified upon the validity of the ergodic hypothesis, which requires that 
the whole phase space should be equally visited in the infinite-time limit.
Typically, this occurs for large Hamiltonian systems, with dynamical 
variables connected by short-range interactions, leading to a microscopic 
dynamics characterized by a quasi-continuum Lyapunov spectrum, whose largest 
value remains
positive in the thermodynamic limit. Such a chaotic dynamical behavior leads 
to a quick occupation of phase space, ensuring a safe use of the ensemble 
theory, and so, the BG statistics may be connected to the standard 
extensive thermodynamics in the well-known elegant manner. However, ergodicity and,  
consequently, the validity of the standard equilibrium ensembles, 
depend crucially on
the nature of the Hamiltonian system considered \cite{balescu}. In 
particular, systems characterized by long-range interactions, or long-range 
microscopic memory, may present a breakdown of ergodicity, leading to 
a fractal (or even more complex)
structure in phase space. In such cases, the BG statistical-mechanics 
framework looses its validity and some more general theory must be 
employed. Recently, a large variety of evidences has been presented, 
exhibiting results that do not conform with the BG formalism; 
one may mention observations on 
turbulent plasmas \cite{boghosian}, 
turbulent fluids \cite{beck01a,beck01b,solomon}, astrophysical 
systems \cite{lyndenbell,labini,milanovic,koyama,torcini},
quantum chaos \cite{weinstein}, logistic map \cite{borges},
glasses \cite{parisi,benedetti}, and complex systems
\cite{upaddhyaya,viswanathan}, among others. Such systems exhibit  
evident inconsistencies with one of the main characteristics in the BG
formalism, which is the extensivity of the entropy and the exponential
weight factor associated with it. For an $N$-particle
system, the extensivity property means that quantities like the internal
or free energy {\it per particle}
should approach a well-defined finite thermodynamic limit when $N \rightarrow \infty $. When 
long-range interactions are present, each microscopic constituent of 
the system interacts with all the others, leading to an energy 
that depends more than linearly with $N$, and, obviously, to a nonextensive 
behavior. As a consequence of this, 
additivity does not hold, in the sense that the free energy of the 
whole system is not equal to the sum of the free energies of its 
macroscopic parts, and the application of the BG formalism becomes 
questionable. A generalized nonextensive thermostatistical formalism, 
proposed over a decade ago \cite{nesmformalism,reviews},
seems to be a good candidate to deal with such systems.

Recently, a lot of attention has been dedicated to a classical
Hamiltonian system,
namely, the inertial long-range-interaction XY model, which consists 
in an assembly of $N$ classical planar rotators interacting 
through a long-range potential 
\cite{antoni,latora98,anteneodo,latora99a,latora99b,latora00a,%
latora00b,latora01a,latora01b,tamarit,campa00,campa01,latora02,bene}. Such a 
system, which has been investigated numerically within the microcanonical 
ensemble, presented clear indications that a more general 
statistical-mechanics formalism is required for its description. 
In particular, for the case of infinite-range interactions, i.e., 
mean-field limit, for which a well-known continuous phase transition 
occurs, if one considers a
total energy close to and below the critical energy, there exists a basin
of attraction for the initial conditions for which the system gets
captured in a metastable state, whose duration increases with $N$, before 
attaining the terminal thermal equilibrium. 
Therefore, if one considers the 
thermodynamic limit ($N \rightarrow \infty$) {\it before} the long-time 
limit, the system will remain in the metastable state and will
never reach the terminal equilibrium state, in such a way that  
the phase space will {\it not} be equally and completely covered. 
Moreover, in such a metastable state,  
the maximum Lyapunov exponent approaches zero, as 
$N \rightarrow \infty$, contrary to what is expected in a standard BG 
equilibrium state.  

In the present work we perform molecular dynamical investigations,
of the isolated inertial
infinite-range-interaction Heisenberg ferromagnet, defined by $N$ classical 
Heisenberg-like rotators. We show that the metastable state that 
occurs in the corresponding XY version of the model, near criticality, 
now appears in a noticeable way for a much wider extent, 
for an energy range that starts right below criticality and 
prolongs up to very
high energies. In the next section, we review the
equilibrium canonical-ensemble solution of the model. In section 3 we 
present the results of our numerical investigation, and finally, in 
section 4, we present our main conclusions. 

\newpage

\noindent
{\large\bf 2. \quad The Equilibrium Canonical-Ensemble Solution}

\vskip \baselineskip

Let us now work out the equilibrium 
canonical-ensemble solution of the model. For the sake of 
generality, throughout this section, we will deal with an inertial
$n$-vector ferromagnet, 
composed by $N$ classical rotators, each of them defined in an 
$n$-dimensional configurational space. 
The Hamiltonian of the system is given by

\vspace{5mm}

\setcounter{enumi}{2}
\setcounter{enumii}{1}
\renewcommand{\theequation}{\arabic{enumi}.\arabic{enumii}}
\begin{eqnarray}
H & = & {1 \over 2} \sum_{i=1}^{N} \sum_{\mu=1}^{n} L_{i \mu}^{2}
+ {1 \over 2N} \sum_{i,j=1}^{N} (1 - \vec{S}_{i}.\vec{S}_{j})
\nonumber \\
& = & {1 \over 2} \sum_{i=1}^{N} \sum_{\mu=1}^{n} L_{i \mu}^{2}
+ {1 \over 2N} \sum_{i,j=1}^{N} \left(1 - \sum_{\mu=1}^{n}
S_{i \mu} S_{j \mu} \right), 
\end{eqnarray}

\vspace{5mm}

\noindent
where the index $\mu$ ($\mu = 1,2, \cdots , n$) denotes Cartesian 
components and $L_{i \mu}$ represents the $\mu$-component of the 
angular momentum (or the rotational velocity, since we are assuming 
unit inertial moments) of rotator $i$.
It is important to remind that the $N$-dependence in the coupling
constant above, is usually (but not necessarily, see \cite{anteneodo}) 
introduced in order to yield a sensible thermodynamic
limit, i.e, a finite free energy per particle when $N \rightarrow \infty$,
within the equilibrium ensemble theory of standard statistical mechanics.

The rotators are allowed to vary
their directions continuously inside an 
$n$-dimensional sphere of unit radius, leading to the constraint

$$
\sum_{\mu=1}^{n} S_{i \mu}^{2} = 1 \qquad (i=1,2, \cdots N).
\eqno(2.2)
$$

\vskip \baselineskip
\noindent
It should be mentioned that such a constraint reduces the 
number of degrees of freedom per particle to $n-1$, in such a way 
that the total number of degrees of freedom of the system 
is given by $N(n-1)$. The model defined above recovers, as particular 
cases, the mean-field inertial XY (whose equilibrium 
canonical-ensemble solution was presented in Ref. \cite{antoni}) and 
Heisenberg (whose dynamics will be discussed in the next section) 
models, for $n=2$ and $n=3$, respectively. Although our model is composed 
by one-dimensional inertial constituents (rotators), we shall 
sometimes refer to $\vec{S} \equiv (S_{1},S_{2}, \cdots , S_{n})$ as 
spin variables, considering the close analogy of the above-defined 
model with the standard $n$-vector ferromagnet.  

One may now follow the standard procedure by rewriting the 
Hamiltonian as

$$
H = {N \over 2} + {1 \over 2} \sum_{i=1}^{N} \sum_{\mu=1}^{n} L_{i \mu}^{2}
- {1 \over 2N} \sum_{\mu=1}^{n} 
\left( \sum_{i=1}^{N}S_{i \mu} \right)^{2}, 
\eqno(2.3)
$$

\vskip \baselineskip
\noindent
in such a way that the partition function becomes

\vspace{5mm}

\setcounter{enumi}{2}
\setcounter{enumii}{4}
\renewcommand{\theequation}{\arabic{enumi}.\arabic{enumii}}
\begin{eqnarray}
Z & = & {\rm exp}\left( {- \beta N \over 2} \right)
\int \left( \prod_{i=1}^{N} \prod_{\mu=1}^{n}dS_{i \mu} \right)
\left( \prod_{i=1}^{N} \prod_{\mu=1}^{n-1}dL_{i \mu} \right)
\left[ \prod_{i=1}^{N} \delta \left( 
\sum_{\mu=1}^{n} S_{i \mu}^{2} - 1 \right) \right]
\nonumber \\
& \times & \left[ \prod_{i=1}^{N} \prod_{\mu=1}^{n-1} {\rm exp}
\left( {-\beta L_{i \mu}^{2} \over 2} \right) \right]
\left\{ \prod_{\mu=1}^{n} {\rm exp} \left[ \left( {\beta \over 2N} 
\right) \left( \sum_{i=1}^{N}S_{i \mu}\right)^{2} \right] \right\}.
\end{eqnarray}

\vspace{5mm}

\noindent
The constraints of Eq. (2.2) are taken into account in the equation 
above, through the delta functions, as well as in the products over 
the angular-momentum variables, which apply only over the effective 
number of degrees of freedom per particle, 
$\mu=1,2 \cdots , n-1$. The squared $\sum_{i}$ may be linearized through 
the application of a Hubbard-Stratonovich-like transformation \cite{stanley}, 
which introduces a set of parameters $\{ x_{\mu} \}$; rescaling 
$\{ x_{\mu} \} \rightarrow \sqrt{N} \{ x_{\mu} \}$ one gets

\vspace{5mm}

\setcounter{enumi}{2}
\setcounter{enumii}{5}
\renewcommand{\theequation}{\arabic{enumi}.\arabic{enumii}}
\begin{eqnarray}
Z & = & \left( {2 \pi \over \beta}\right)^{N(n-1)/2} 
{\rm exp}\left( {- \beta N \over 2} \right)
\int \left\{ \prod_{\mu=1}^{n} \left[ 
\left( {N \over 2\pi} \right)^{1/2} dx_{\mu} 
{\rm exp} \left( {-Nx_{\mu}^{2} \over 2} \right) \right] \right\}
\nonumber \\
& \times & \prod_{i=1}^{N} \left\{ \int \prod_{\mu=1}^{n} \left[ 
dS_{i \mu} {\rm exp}
\left( \beta^{1/2} S_{i \mu}x_{\mu} \right) \right]
\delta \left( \sum_{\mu=1}^{n} S_{i \mu}^{2} - 1 \right) \right\}.
\end{eqnarray}

\vspace{5mm}

\noindent
As usual, the site index may be discarded, and straightforward 
calculations lead to

\vspace{5mm}

\setcounter{enumi}{2}
\setcounter{enumii}{6}
\renewcommand{\theequation}{\arabic{enumi}.\arabic{enumii}}
\begin{eqnarray}
Z & = & \left( {2 \pi \over \beta}\right)^{N(n-1)/2} 
{\rm exp}\left( {- \beta N \over 2} \right)
\int \left[ \prod_{\mu=1}^{n}  
\left( {N \over 2\pi} \right)^{1/2} dx_{\mu} \right]
{\rm exp} \left[ -N \left( \sum_{\mu=1}^{n} {x_{\mu}^{2} \over 2}
- \ln \xi \right) \right],
\end{eqnarray}

\vspace{5mm}

\noindent
where

$$
\xi = 2^{(n-2)/2} \ \pi^{n} y^{-(n-2)/2}
I_{(n-2)/2} (y); 
\qquad y = \left( \beta \sum_{\mu=1}^{n} x_{\mu}^{2} \right)^{1/2},
\eqno(2.7)
$$

\vskip \baselineskip
\noindent
with $I_{k}(y)$ denoting modified Bessel functions of the first kind 
of order $k$. Considering $N$ large, one may use steepest descents,

$$
Z \approx \left( {2 \pi \over \beta}\right)^{N(n-1)/2} 
{\rm exp}\left( {- \beta N \over 2} \right)
{\rm exp} \left[ -N {\rm max}_{y} \left(
{y^{2} \over 2\beta} - \ln \xi (y) \right) \right],
\eqno(2.8)
$$

\vskip \baselineskip
\noindent
with the condition of maximum leading to the self-consistent equation,

$$
\bar{y} = \beta \left( {1 \over \xi} {d\xi \over dy} \right)_{y=\bar{y}}
= \beta {I_{n/2}(\bar{y}) \over I_{(n-2)/2}(\bar{y})}.
\eqno(2.9)
$$

\vskip \baselineskip
\noindent
In the thermodynamic limit, one obtains the free-energy per particle, 

$$
\beta f = {\beta \over 2} - {n-1 \over 2} \ln 
\left( {2\pi \over \beta} \right)
+{\bar{y}^{2} \over 2\beta} - \ln \xi (\bar{y}),
\eqno(2.10)
$$

\vskip \baselineskip
\noindent
as well as the internal energy per particle,

$$
u = {n-1 \over 2 \beta} + {1 \over 2} (1 - \vec{m}^{2}),
\eqno(2.11)
$$

\vskip \baselineskip
\noindent
where $\vec{m}$ represents the magnetization per particle, whose modulus
is directly related to the parameter $\bar{y}$,

$$
m \equiv |\vec{m}| = {\bar{y} \over \beta}
= {I_{n/2}(\beta m) \over I_{(n-2)/2}(\beta m)}.
\eqno(2.12)
$$

\vskip \baselineskip
\noindent
The critical temperature of the model may be obtained by considering $m$ 
small and expanding the r.h.s. of Eq. (2.9) in power series 
(we work in units of $k_{B}=1$), 

$$
T_{c} = {1 \over 2} {\Gamma (n/2) \over \Gamma [(n+2)/2]} = {1 \over n},
\eqno(2.13)
$$

\vskip \baselineskip
\noindent
which may be substituted into Eq. (2.11) to yield the critical energy 
density,

$$
u_{c} = {1 \over 2} [1 + (n-1)T_{c}] = 1 - {1 \over 2n}
\eqno(2.14)
$$

\vskip \baselineskip
\noindent
The above results recover those obtainted in Ref. \cite{antoni}, 
for $n=2$, e.g., the self-consistent equation,

$$
\bar{y} = \beta {I_{1}(\bar{y}) \over I_{0}(\bar{y})},
\eqno(2.15)
$$

\vskip \baselineskip
\noindent
leading to $T_{c}=1/2$ and $u_{c}=3/4$. For the case $n=3$, one gets 
the well-known self-consistent equation,

$$
m = {\rm cotanh}(\beta m) - {1 \over \beta m},
\eqno(2.16)
$$

\vskip \baselineskip
\noindent
with $T_{c}=1/3$ and $u_{c}=5/6$.
For $n$ increasing from unit (Ising model) to infinity (spherical model), 
$T_{c}$ decreases form 1 to 0, and $u_{c}$ increases from $1/2$ to 1. 

In the next section we discuss the dynamics of the particular 
case $n=3$ of the model defined above. 

\vskip 2\baselineskip

\noindent
{\large\bf 3. \quad Dynamics of the Mean-Field Inertial Heisenberg 
Model}

\vskip \baselineskip

\noindent
{\bf 3.1 - Molecular Dynamics \\ }
In this section we will present the results obtained by 
simulations of the constant-energy dynamics of the model defined by 
Eqs. (2.1) and (2.2) for the particular case $n=3$, i.e., Heisenberg-like 
rotators. From now on, we will use the standard notation for 
the Cartesian components of a Heisenberg model, i.e., 
$\mu=x,y,z$. The results to be discussed below were obtained by a direct 
integration of the equations of motion,

\vspace{5mm}

\setcounter{enumi}{3}
\setcounter{enumii}{1}
\setcounter{equation}{0}
\renewcommand{\theequation}{\arabic{enumi}.\arabic{enumii}\alph{equation}}
\begin{eqnarray}
\dot{\vec{L}}_{i} & = & \vec{S}_{i} \times \left( {1 \over N} 
\sum_{j=1}^{N} \vec{S}_{j} \right) \qquad (i=1,2, \cdots , N),  \\
\dot{\vec{S}}_{i} & = & \vec{L}_{i} \times \vec{S}_{i}
\qquad (i=1,2, \cdots , N), 
\end{eqnarray}

\vspace{5mm}

\noindent
which correspond to a set of $6N$ equations to be solved numerically.
  
For solving such a set of equations we have used a fourth-order 
Runge-Kutta-Merson integrator 
\cite{lapidus} with a time step of 0.05, leading, respectively,
to the relative energy
and spin-normalization conservations of $10^{-4}$ and $10^{-3}$, 
or better. The total initial kinetic energy was divided
into three equal parts, each of them to be assigned 
to a given set of Cartesian components of angular velocities
$\{ L_{i \mu} \} \ (i=1,2, \cdots , N)$. We have always 
started the system with 
the so-called water-bag initial conditions 
\cite{latora01a,latora01b,latora02} 
for each set of components of angular velocities, i.e.,  
each set $\{ L_{i \mu} \}$ was extracted from a symmetric uniform distribution 
and then, 
translated and rescaled to have zero total momentum. In what concerns 
the spin variables, we have started our simulations with all spins 
aligned along the $z$-axis (zero initial potential energy).
Our measured quantities correspond to averages over $N_{s}$ distinct samples,
i.e., different initial sets of $\{ L_{i \mu} \}$.

\vskip \baselineskip

\noindent
{\bf 3.2 - Caloric Curve and Metastability \\ }

In Fig. 1 we exhibit the caloric curve (full line) obtained by solving
the equilibrium canonical-ensemble equations [Eqs. (2.11) and (2.12)]. 
We have chosen four particular values of the energy density to
investigate, within our microcanonical-ensemble molecular dynamical
approach,
how $\langle K \rangle /N$ evolves in 
time, for different values of $N$ (it should be mentioned that, the
quantity $\langle K \rangle /N$, when evaluated at 
equilibrium, is expected to coincide with the temperature).
Two of the chosen energies, $u=0.75$ [Fig. 2(a)] and $u=0.96$ [Fig. 2(b)],
correspond, respectively, to values slightly below and above the critical
internal energy ($u_{c}=5/6$).
The energy $u=1.32$ [Fig. 2(c)] is inside a range of energies where the
kinetic temperature $\langle K \rangle /N$ presents a maximum discrepancy
between its values at intermediate and long times. Finally, the fourth chosen
energy, $u=7.2$ [Fig. 2(d)] corresponds to a value far above $u_{c}$.
In all the plots of Figs.
2(a)--(d) the system was started with the above-mentioned initial conditions
and we have considered $N_{s}=16$ ($N=200$), $N_{s}=12$ ($N=400$),
$N_{s}=8$ ($N=800$), and $N_{s}=4$ ($N=1600$).
We have observed that, after a short transient, the system
rapidly reached a metastable,
or quasi-stationary state (QSS), with a value
of $\langle K \rangle /N$ {\it higher} than the one predicted by the
canonical-ensemble equilibrium theory. It is important to remind that a 
QSS has been found also for the corresponding XY version of 
the present model, in an unambiguous way, only near criticality 
\cite{latora98,latora99a,latora00a,latora01a,latora01b,latora02}. Except 
for very low energies, the QSS is always detected easily in the
Heisenberg case;
this is shown in Fig. 1, where we exhibit the values of
$\langle K \rangle /N$ for both  
QSSs (empty squares) and terminal equilibrium states (empty circles), for 
systems with $N=400$ and different values of the energy density. 
One clearly sees that, in what concerns the value of 
$\langle K \rangle /N$ at the terminal equilibrium 
state, the thermodynamic limit may be attained within our computational effort
(in the sense that the
present microcanonical-ensemble numerical approach agrees with the 
equilibrium canonical-ensemble results) for $u \gg u_{c}$, 
whereas for $u \sim u_{c}$, one observes strong finite-size effects. In 
fact, the terminal-equilibrium value of $\langle K \rangle /N$ 
seems to reach its thermodynamic limit for very small systems, if 
$u \gg u_{c}$ [as shown in Fig. 2(d), for the case $u=7.2$, where
the size $N=200$ appears to have converged], 
whereas near criticality ($u \sim u_{c}$) one may observe
fluctuations [as shown in Figs. 2(a) and 2(b)].

Although large fluctuations may be observed on the values of 
$\langle K \rangle /N$ in 
the QSS, it appears evident that the gap with respect to the 
corresponding terminal-equilibrium-state values survives in 
the thermodynamic limit. If one defines the 
lifetime of the QSS ($t_{\rm QSS}$) as the time at which 
$\langle K \rangle /N$ presents 
its half-way between the values at the QSS and the terminal thermal 
equilibrium, one concludes that such a quantity increases, essentially, 
linearly with $N$, as shown in Fig. 3, for the energies considered in
Figs. 2(a)--(d). Therefore, the duration of the QSS increases with $N$, in
such a way that, if the thermodynamic limit is performed {\it before} the 
long-time limit, the system will never relax to the terminal thermal 
equilibrium.

\newpage

\noindent
{\bf 3.3 - Sensitivity to the Initial Conditions \\ }
Let us now investigate how the maximal Lyapunov exponent,
$\lambda_{\rm max}$ scales with $N$. 
Herein we shall use the well-known 
method for calculating such a quantity by considering the limit \cite{benettin},

\vspace{5mm}

\setcounter{enumi}{3}
\setcounter{enumii}{2}
\setcounter{equation}{0}
\renewcommand{\theequation}{\arabic{enumi}.\arabic{enumii}\alph{equation}}
\begin{eqnarray}
\lambda_{\rm max} & = & \lim_{t \rightarrow \infty}{1 \over t} 
\ln \left[ {d(t) \over d(0)} \right] 
= \lim_{t \rightarrow \infty} \lambda (t), \\
d(t) & = & \left\{ \sum_{i=1}^{N} \sum_{\mu=x,y,z} \left[ 
(\delta L_{i \mu})^{2} + (\delta S_{i \mu})^{2} \right] \right\}^{1/2}, 
\end{eqnarray}

\vspace{5mm}

\noindent
where $d(t)$ represents the metric distance calculated from infinitesimal 
displacements in phase space, at time $t$. We have carried simulations
up to $t=10000$ (i.e., 200000 time steps), for different system sizes 
($N=50,100,400,1600,3200$), and $\lambda_{\rm max}$ was obtained after
averaging over $N_{s}=50$ samples for the smallest size ($N=50$), whereas
for all other sizes we have considered $N_{s}=10$. For
the cases where there is an apparent QSS, the time interval considered
ensures that the
$\lambda_{\rm max}$ computed does indeed correspond to a quantity 
in the QSS [see, e.g., Figs. 2(a)--(d)],
whereas for the cases where there is no evident QSS, the time used
is expected to be sufficient for the system to have reached 
its terminal thermal equilibrium. 
As shown in Fig. 4, one has that $\lambda_{\rm max} \sim N^{-\kappa}$,
similarly to what happens for the XY 
version of the present model \cite{latora98,latora01b,latora02,bene}. We 
computed $\kappa$ for energies below $u_{c}$ ($u=0.30$), as well as for 
$u \gg u_{c}$ ($u=7.2$), with the above-mentioned initial conditions; we 
have obtained essentially the same estimate in both cases, 
$\kappa=0.225 \pm 0.030$. It is important to remind that, in the mean-field 
inertial XY model, one has $\kappa=1/3$ for $u \gg u_{c}$
\cite{latora98,latora01b,latora02}, whereas $\kappa=1/9$ for 
$u<u_{c}$ \cite{bene}; in the former case, there is no apparent QSS, in
such a way that the estimate $\kappa=1/3$ is expected to apply to a terminal
thermal equilibrium state,
whereas in the latter, $\kappa$ was obtained at the peculiar QSS. Such 
estimates find no similar with those presented herein for the Heisenberg 
case, which apply to the QSS for $u \gg u_{c}$, whereas for the low energy 
considered ($u=0.30$) in the case $u < u_{c}$ we have found no clear evidence of a QSS 
(although the presence of a QSS, with a small gap, indiscernible due to 
fluctuations in $\langle K \rangle /N$, with respect to the 
terminal equilibrium state, is not ruled out). However, the common feature 
observed for both XY and Heisenberg models, 
$\lambda_{\rm max} \sim N^{-\kappa}$ (yielding a zero maximal Lyapunov 
exponent in the thermodynamic limit), which seems to hold in the presence or not 
of an evident QSS, does certainly contradict the standard BG theory 
for an equilibrium state (which requires a finite $\lambda_{\rm max}$ in 
the thermodynamic limit). 

\vskip 2\baselineskip

\noindent
{\large\bf 4. \quad Conclusion}

\vskip \baselineskip

We have analysed a system of $N$ Heisenberg-like classical rotators with 
ferromagnetic infinite-range interactions. The dynamics of the model was 
studied within the microcanonical ensemble, by directly solving the 
equations of motion. For a finite $N$, the time evolution of the kinetic 
temperature shows that there is a basin of attraction for the
initial conditions for which the
system gets caught in a metastable state, before reaching the terminal
thermal equilibrium. We have shown that the duration of such a metastable 
state diverges as $N \rightarrow \infty$, tipically linearly with $N$. 
Therefore, if the thermodynamic limit is considered before the long-time 
limit, the system will never relax to the terminal thermal equilibrium. We 
have also calculated the maximum Lyapunov exponent, above and below the 
critical point; in both cases, the scaling,  
$\lambda_{\rm max} \sim N^{-\kappa}$, was verified. For the particular 
initial conditions considered (maximal magnetization), the exponent $\kappa$ presents the same 
value for energies chosen above and below the critical point,
$\kappa=0.225 \pm 0.030$. Above the critical point our estimate applies to
the metastable state, whereas the estimate below the critical point is
expected to hold for the terminal thermal equilibrium, since in such a case
we have found no clear evidence of the existence of a metastable state.
Preliminary
studies suggest that, above criticality, the exponent $\kappa$ does not 
depend, within the error bars, on the initial conditions employed; however, 
below criticality, different initial conditions may possibly lead to a 
breakdown of universality, with different estimates for $\kappa$.
In particular, the $\kappa$ estimates below criticality seem to vary if
the initial conditions for the spin variables break or not the
Heisenberg-like symmetry of the system, i.e., if we start with $m \ne 0$ (present 
paper) or $m = 0$. 
Further studies to clarify this point constitute the next step along the present 
lines. 

\vskip 2\baselineskip

{\large\bf Acknowledgments}

\vskip \baselineskip
\noindent
We thank fruitful conversations with C. Anteneodo, E.~P. Borges, and 
E.~M.~F. Curado. 
The partial financial supports from
CNPq, Pronex/MCT and FAPERJ (Brazilian agencies) are acknowledged.
One of us (F.~D.~N.)
acknowledges CBPF (Centro Brasileiro de Pesquisas F\'{\i}sicas) for the 
warm
hospitality during a visiting period in which this work was accomplished.

\newpage

\centerline{{\large\bf Figure Captions}}

\vskip 2\baselineskip
\noindent
{\bf Fig. 1:} \qquad The caloric curve for the inertial mean-field
ferromagnetic Heisenberg model obtained by the equilibrium 
canonical-ensemble solution (full line). In the vertical axis, 
$\langle K \rangle /N$, is a quantity that is expected to coincide with 
the temperature, at equilibrium. The dashed vertical line signals 
the second-order phase transition critical energy density, $u_{c}=5/6$. The 
empty squares and circles represent the 
estimates of $\langle K \rangle /N$ from the microcanonical-ensemble 
numerical analysis of a system of size $N=400$, 
at the metastable state and after that, respectively.  

\vskip 2\baselineskip
\noindent
{\bf Fig. 2:} \qquad The microcanonical time evolution of
$\langle K \rangle /N$ is represented
for several system sizes and different energy densities: just below
criticality [$u=0.75$ (a)], just above
criticality [$u=0.96$ (b)], at the region where the gap between the 
metastable and terminal equilibrium states is maximum [$u=1.32$ (c)] and 
for $u \gg u_{c}$ [$u=7.2$ (d)].
The initial conditions are water-bag for velocities and $m=1$. 

\vskip 2\baselineskip
\noindent
{\bf Fig. 3:} \qquad Log-log plots of the lifetime 
($t_{\rm QSS}$) of the QSS as a function of $N$, for the energy densities 
considered 
in Figs. 2(a)--(d).
Simple linear fits yield the slopes $0.99 \pm 0.05$ ($u=0.75$),
$1.10 \pm 0.06$ ($u=0.96$),
$1.01 \pm 0.02$ ($u=1.32$), and $0.95 \pm 0.06$ ($u=7.20$). In each case, 
the slope is very close to 1 (represented by the dashed line), in such a 
way that $t_{\rm QSS} \sim N$ in all cases. 
The initial conditions are water-bag for velocities and $m=1$. 

\vskip 2\baselineskip
\noindent
{\bf Fig. 4:} \qquad Log-log plots of the maximum Lyapunov exponent
$\lambda_{\rm max}$ versus $N$, for energies densities below ($u=0.3$) 
and above the critical point ($u=7.2$), showing the scaling 
$\lambda_{\rm max} \sim N^{-\kappa}$. The slopes are essentially the same, 
within the error bars, yielding $\kappa=0.225 \pm 0.030$. For $u=7.2$, the 
distinction between the metastable and teminal equilibrium states is very clear, 
and this value of $\kappa$ corresponds to the metastable one. For $u=0.3$, the 
distinction in not very clear, and this value of $\kappa$ presumably corresponds 
to the terminal equilibrium state.

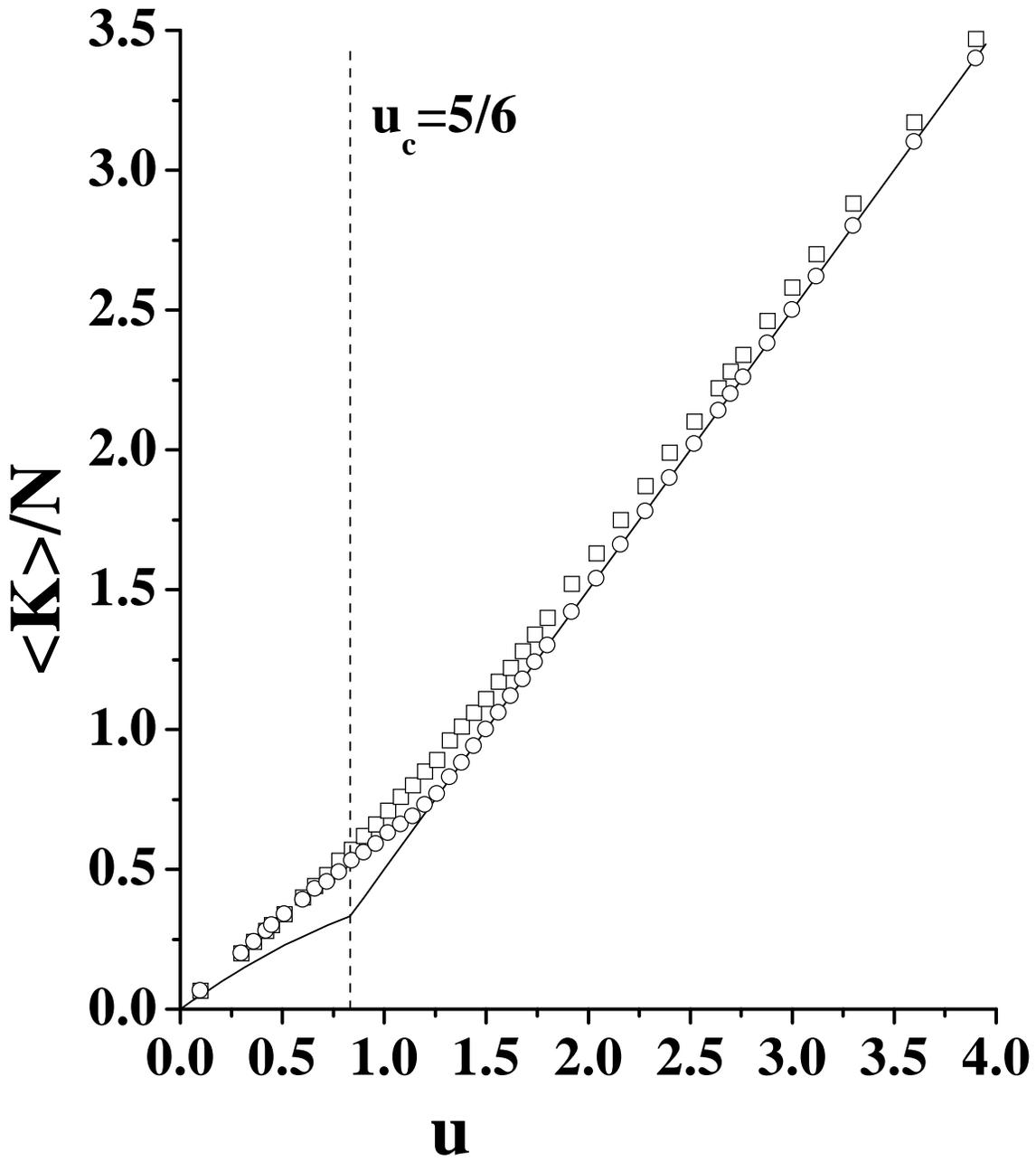
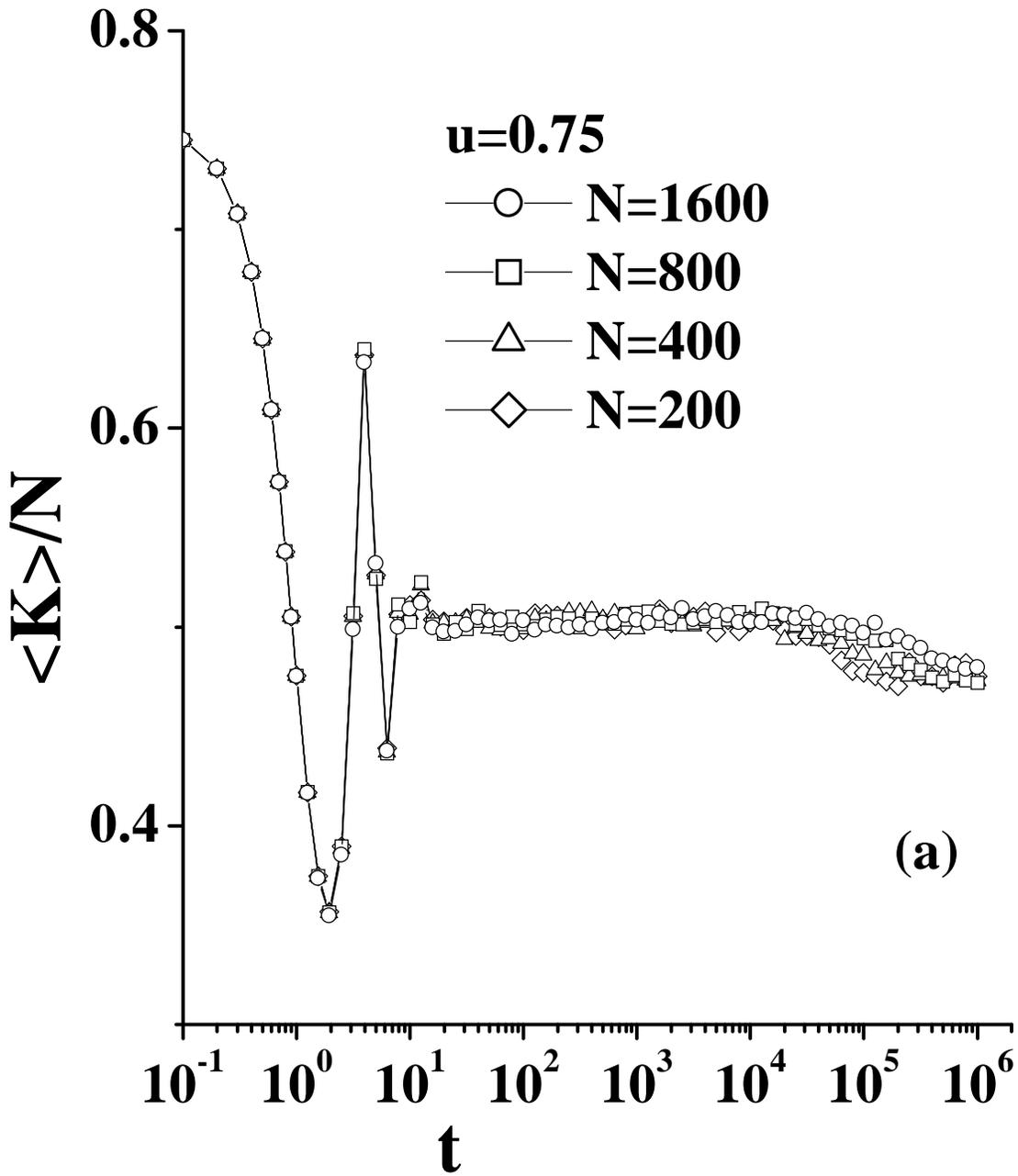
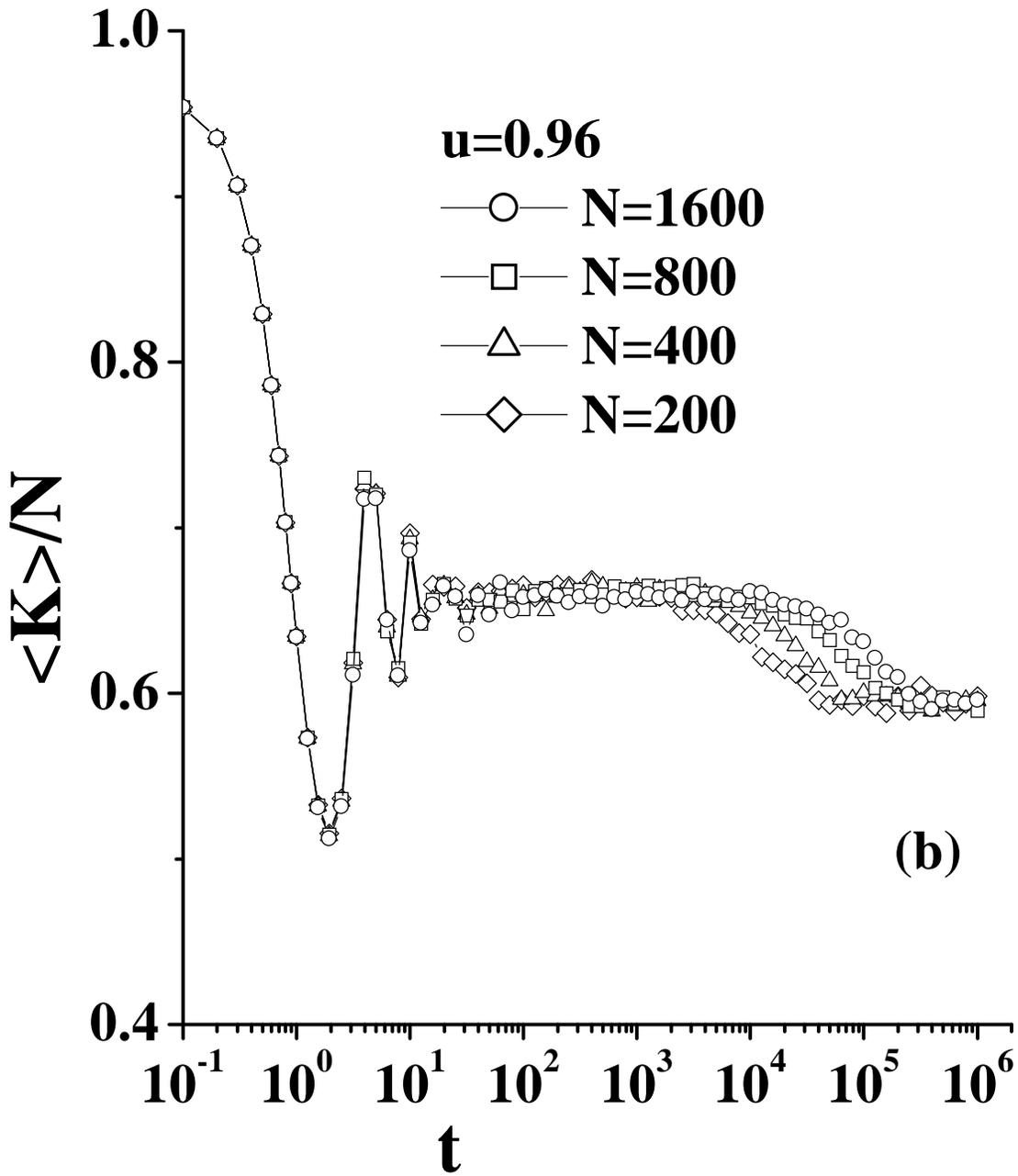
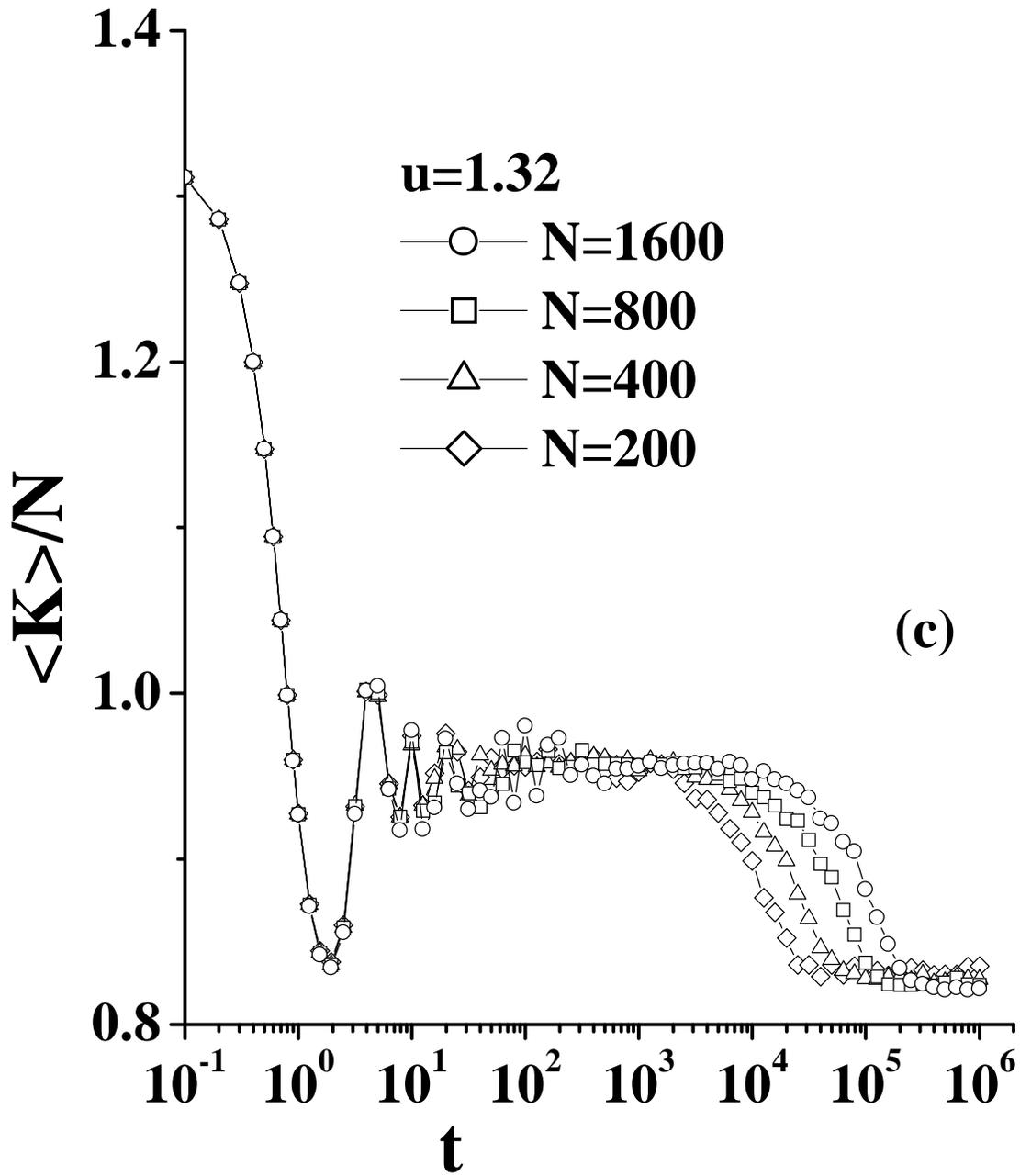
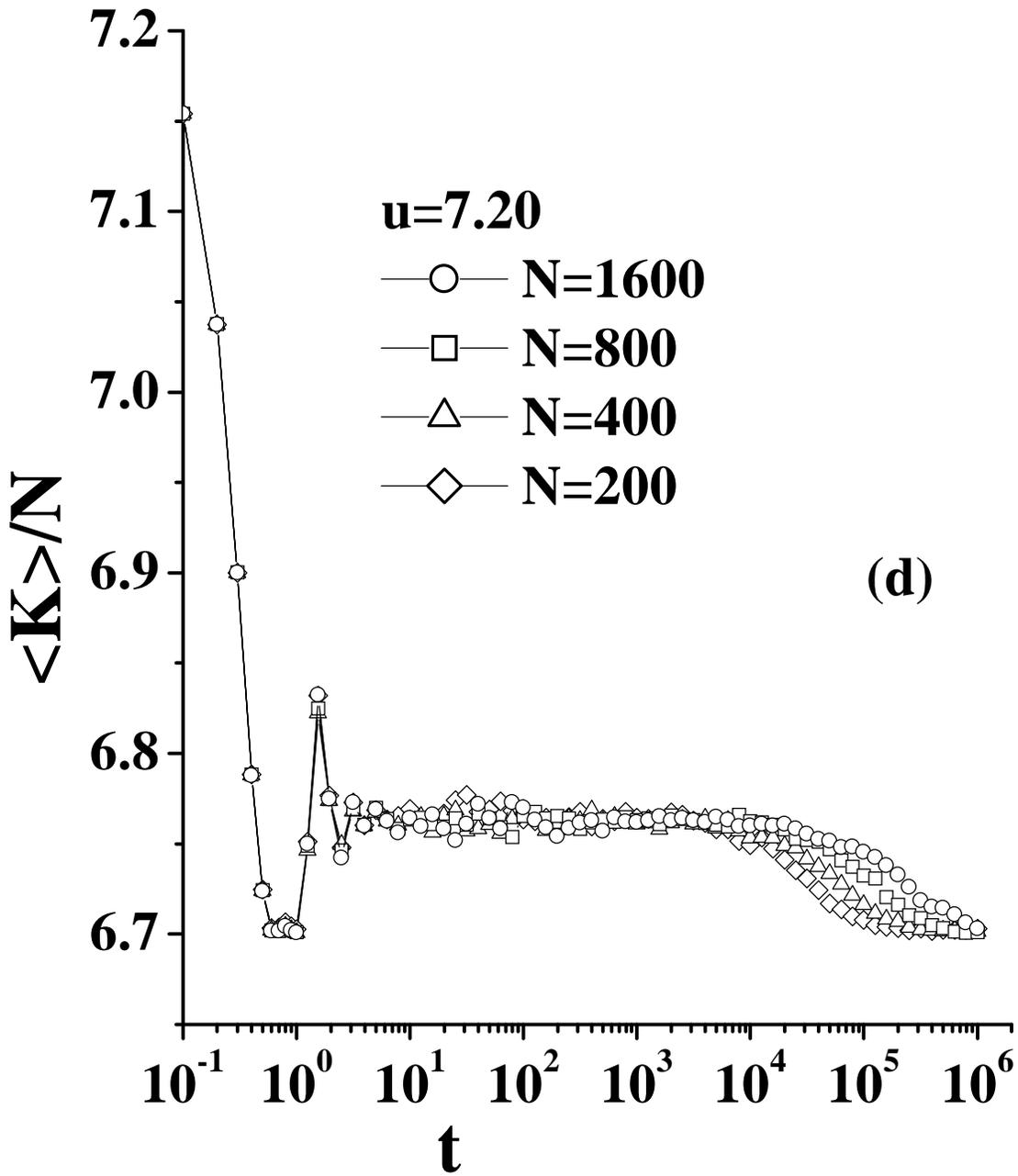
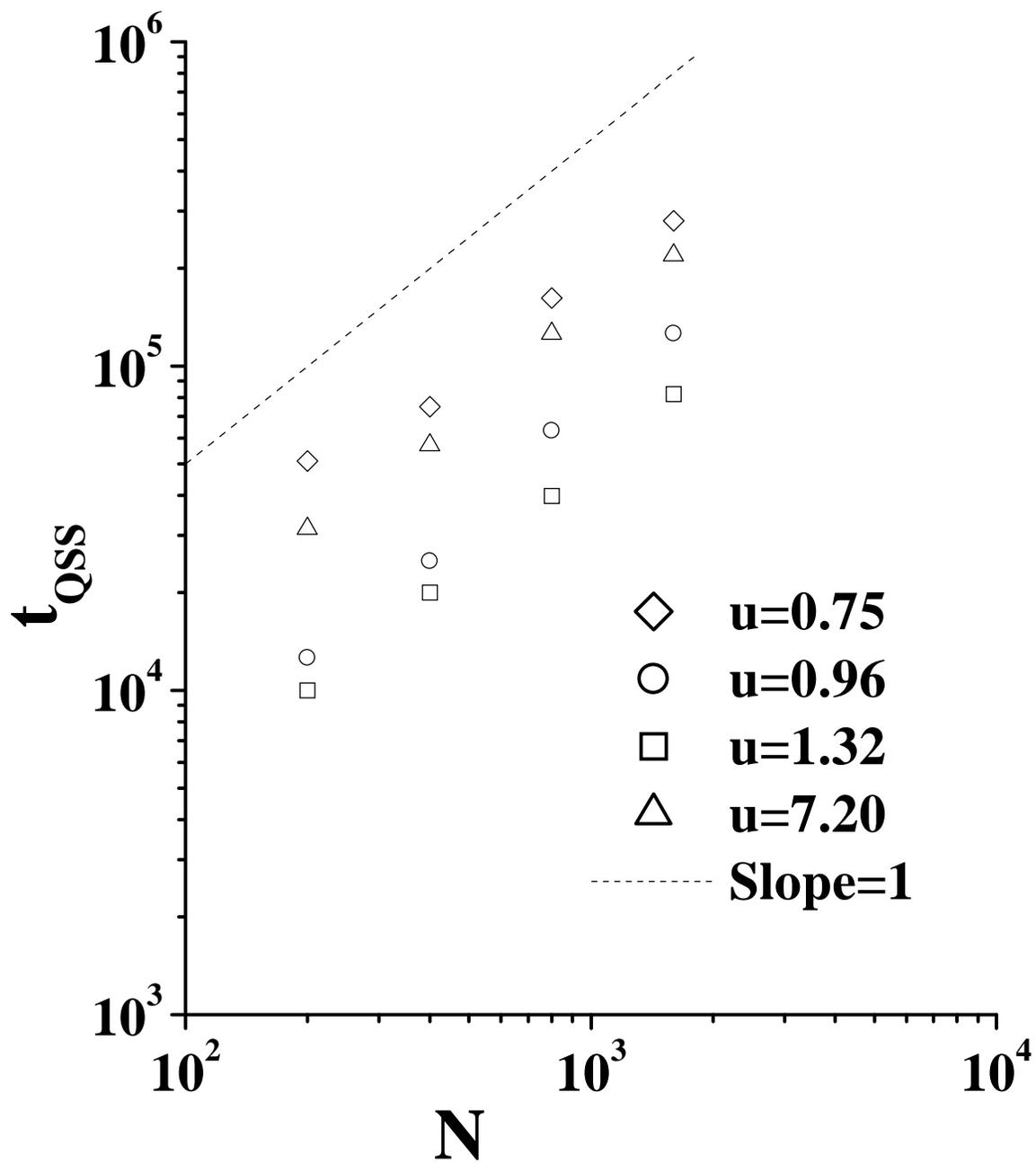
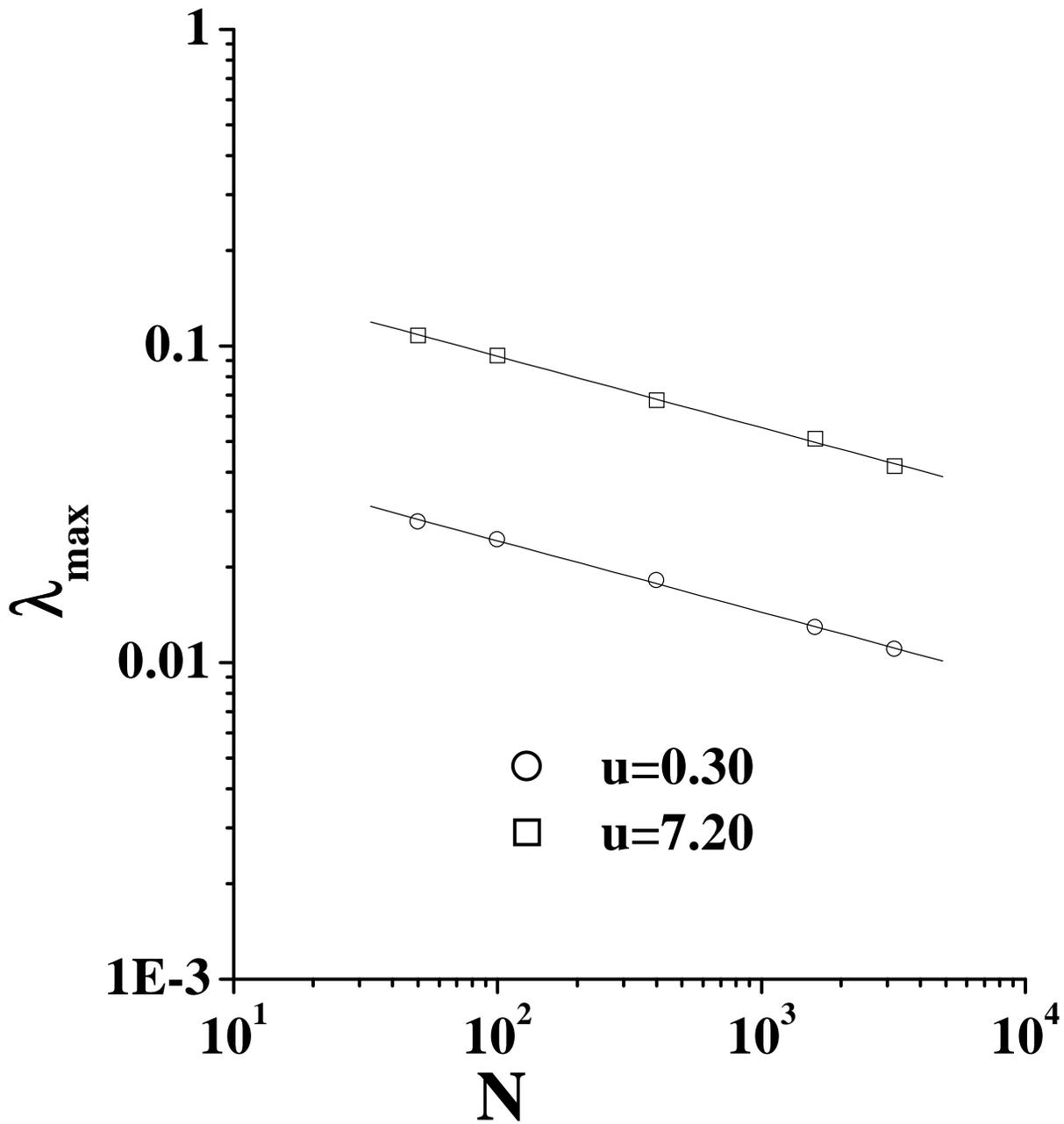
\end{document}